\documentstyle[here,epsfig]{mn2e}
\def\simlt{\mathrel{\rlap{\lower 3pt\hbox{$\sim$}}\raise 2.0pt\hbox{$<$}}}
\def\simgt{\mathrel{\rlap{\lower 3pt\hbox{$\sim$}} \raise 2.0pt\hbox{$>$}}}

\def\gtsima{$\; \buildrel > \over \sim \;$}
\def\ltsima{$\; \buildrel < \over \sim \;$}
\def\gtrsim{\lower.5ex\hbox{\gtsima}}
\def\lesssim{\lower.5ex\hbox{\ltsima}}
\def\url#1{{\ttfamily\def\/{/\diskretionary{}{}{}}#1}}

\newcommand{\q}{\begin{equation}}
\newcommand{\qa}{\begin{eqnarray}}
\newcommand{\qs}{\begin{eqnarray*}}
\newcommand{\nq}{\end{equation}}
\newcommand{\nqa}{\end{eqnarray}}
\newcommand{\nqs}{\end{eqnarray*}}

\begin{document}

\title[Are ring galaxies the ancestors of GLSBs?]{Are ring galaxies the ancestors of giant low surface brightness galaxies?}
\author[Mapelli et al.]
{M. Mapelli$^{1}$, B. Moore$^{1}$, E. Ripamonti$^{2,3}$, L. Mayer$^{1,4}$, M. Colpi$^{2}$, L. Giordano$^{2}$
\\
$^{1}$ Institute for Theoretical Physics, University of Z\"urich, Winterthurerstrasse 190, CH-8057, Z\"urich, Switzerland; {\tt mapelli@physik.unizh.ch}\\
$^{2}$ Universit\`a Milano Bicocca, Dipartimento di Fisica G.~Occhialini, Piazza delle Scienze 3, I-20126, Milano, Italy\\
$^{3}$ Kapteyn Astronomical Institute, University of Groningen, Postbus 800, 9747 AD Groningen, the Netherlands\\
$^{4}$ Institute of Astronomy, ETH Z\"urich, ETH Honggerberg HPF D6, CH-8093, Z\"urich, Switzerland\\
}

\maketitle \vspace {7cm }

\begin{abstract}
We simulate the collisional formation of a  ring galaxy and we integrate its evolution up to 1.5 Gyr after the interaction. About $100-200$ Myr after the collision, the simulated galaxy is very similar to observed ring galaxies (e.g. Cartwheel). After this stage, the ring keeps expanding and fades. 
Approximately $0.5-1$ Gyr after the interaction, the disc becomes very large ($\sim{}100$ kpc) and flat. Such extended discs have been observed only in giant low surface brightness galaxies (GLSBs).
We compare various properties of our simulated galaxies (surface brightness profile, morphology, HI spectrum and rotation curve) with the observations of four well-known GLSBs (UGC6614, Malin 1, Malin 2 and NGC7589). The simulations match quite well the observations, suggesting that ring galaxies could be the progenitors of GLSBs. This result is crucial for the cold dark matter (CDM) model, as it was very difficult, so far, to explain the formation of GLSBs within the CDM scenario.

\end{abstract}

\section{Introduction}
Ring galaxies are one of the most intriguing categories of peculiar galaxies. About 280 galaxies have been classified as ring-like in the  Catalogue of Southern Peculiar Galaxies and Associations (CPGA; Arp \& Madore 1987). They have commonly been divided in two different classes, P-type and O-type ring galaxies (Few \& Madore 1986). The former consists of galaxies where the nucleus is often off-centre and the ring is quite knotty and irregular, while the latter includes  objects where the nucleus is central and the ring is regular. Even if for some objects such classification is ambiguous, the number of P-type and  of O-type ring galaxies are roughly comparable.
 
The differences among the two classes are probably connected with the formation mechanism of such galaxies: for P-type ring galaxies a collisional origin has been proposed (Lynds \& Toomre 1976; Theys \& Spiegel 1976; Appleton \& Struck-Marcell 1987a, 1987b; Hernquist \& Weil 1993; Mihos \& Hernquist 1994;  Appleton \& Struck-Marcell 1996; Struck 1997; Horellou \& Combes 2001), as most of them have at least one nearby companion (Few \& Madore 1986); whereas O-type ring galaxies can be resonant (R)S galaxies (de Vacouleurs 1959).

Simulations of galaxy collisions leading to the formation of P-type ring galaxies show that the ring phase is quite short-lived (Hernquist \& Weil 1993; Mihos \& Hernquist 1994; Horellou \& Combes 2001; Mapelli et al. 2007, hereafter M07): the simulated disc galaxy develops a ring similar to the observed ones  $\approx{}100$ Myr after the collision with the intruder; but the ring remains dense and clearly visible only for the first  $\approx{}300$ Myr (M07).

 Thus, ring galaxies are expected to rapidly evolve into something else. Up to now, only  analytic models for ring waves have been adopted to study the late stages of the ring galaxy life (Struck-Marcell \& Lotan 1990;  Appleton \& Struck-Marcell 1996). Neither observations nor simulations have been carried on to investigate  the fate of ring galaxies after the ring phase, leaving a lot of uncertainties. This paper aims at describing the subsequent stages of the evolution of a ring galaxy (up to $\sim{}1.5$ Gyr), by  means of SPH/$N$-body simulations.


Our simulations suggest a possible link between old ring galaxies and the so-called giant low surface brightness galaxies (GLSBs). 
The GLSBs are low surface brightness galaxies (LSBs) characterized by unusually large extension of the stellar and gaseous discs (up to $\sim{}100$ kpc; Bothun et al. 1987; Impey \& Bothun 1989; Bothun et al. 1990; Sprayberry et al. 1995; Pickering et al. 1997; Moore \& Parker 2007) and (often) by the presence of a  normal stellar bulge (Sprayberry et al. 1995; Pickering et al. 1997). 

The existence of GLSBs has always been puzzling. Galaxy formation simulations within the cold dark matter (CDM) model have serious
difficulties in producing realistic disc galaxies. In such simulations, too much angular momentum is
lost during the assembly of objects, producing discs that tend to be too compact and dense.
While a 'Milky Way-like' galaxy can be reproduced by high-resolution CDM simulations (provided that its merging history is 
fairly quiet and that heating by supernovae is properly accounted for; Governato et al. 2007), the extended discs of GLSBs require that much more angular momentum is preserved during the hierarchical build up. Such extended discs are thus beyond the reach of current galaxy formation simulations,
and it is unclear whether improving the realism of such simulations will solve the problem.

Hoffman, Silk \& Wyse (1992) proposed that  GLSBs form from rare density peaks in voids. However, most of currently known GLSBs do not appear to be connected with voids and often show interacting companions (Pickering et al. 1997). A more promising scenario is the formation of GLSBs from  massive disc galaxies due to a bar instability: a large 
scale bar can redistribute the disc matter and significantly increase the disc scale-length (Noguchi 2001; Mayer \& Wadsley 2004).  In this case the redistribution of angular momentum by the
bar instability could counteract the natural tendency of hierarchical assembly to remove angular
momentum from the disc material (Kaufmann et al. 2007). Bar instabilities, however, normally
do not increase the disc scale length by more than a factor of 2-2.5 (Debattista et al. 2006; Kaufmann et al. 2007).
 
In this paper we show that also the propagation of the ring in an old collisional ring galaxy can lead to the redistribution of  mass and angular momentum in both the stellar and gas component
out to a distance of $\sim{}100-150$ kpc from the centre of the galaxy, producing features (e.g. the surface brightness profile, the star formation, the HI emission spectra and the rotation curve) which are typical of GLSBs. 

\section{Simulations}
The method adopted for the simulations analyzed in this paper has already been described in M07 (see also Hernquist 1993; Mapelli, Ferrara \& Rea 2006; Mapelli 2007). Here we briefly summarize the most important points, referring to M07 for the details.

The simulations have been carried out with the parallel SPH/$N-$body code GASOLINE (2005 August version; Wadsley, Stadel \& Quinn 2004) on the cluster {\it zbox}2\footnote{{\tt http://www-theorie.physik.unizh.ch/\~{}dpotter/zbox/}} at the University of Z\"urich. 

The progenitor of the ring galaxy (in the following, 'progenitor' or 'target' galaxy) has been modelled as a disc galaxy, having 4 different components:
\begin{itemize}
\item[(i)] a Navarro, Frenk \& White (1996, hereafter NFW) dark matter halo with  virial mass $M_{vir}=4.9\times{}10^{11}\,{}M_\odot{}$,  virial radius $R_{200}=140$ kpc, and  concentration $c=12$;

\item[(ii)] a stellar exponential Hernquist disc (Hernquist 1993; M07) with mass $M_d$, scale-length $R_d$, and scale-height $z_0$ (see Table 1);

\item[(iii)] a stellar spherical Hernquist bulge (Hernquist 1993; M07) with mass $M_b=2.4\times{}10^{10}\,{}M_\odot{}$, and scale-length $a=0.2\,{}R_d$;

\item[(iv)] a gaseous exponential Hernquist disc (Hernquist 1993; M07) with mass $M_g=3.2\times{}10^{10}\,{}M_\odot{}$, scale-length $R_g=R_d$, and scale-height $z_g=0.057\,{}R_d$ (Hernquist \& Weil 1993). The gas is allowed to cool down to a temperature of $2\times{}10^4$~K, and to form stars according to the recipe described by Katz (1992), whose results agree with the Schmidt-Kennicutt law (Kennicutt 1998, and references therein).

\end{itemize}

The intruder galaxy is a pure NFW dark matter halo with virial mass $M_{vir}=3.2\times{}10^{11}\,{}M_\odot{}$, virial radius $R_{200}=30$ kpc and concentration $c=12$. The initial position and velocity of the centre-of-mass of the intruder (with respect to the centre-of-mass of the progenitor of the ring galaxy) are {\bf x}=(0, 8, 32) kpc and {\bf v}=(28, -218, -872) km s$^{-1}$ (the same as for runs A in M07). This implies that the intruder has very small impact parameter and an inclination angle of $\sim{}14^\circ{}$ with respect to the angular momentum axis of the target. The relative velocity of the intruder is similar to the escape velocity from the target, in order to simulate a flyby. These initial conditions have been chosen in order to simulate the interaction between Cartwheel and its companion G3, accordingly to Horellou \& Combes (2001). In our simulations we adopted a Cartwheel-like system, as Cartwheel is the prototype of collisional ring galaxies.

\subsection{Description of runs}
\begin{table}
\begin{center}
\caption{Initial parameters for the stellar disc.}
\begin{tabular}{cccc}
\hline
\vspace{0.1cm}
Run &  $M_d/(10^{10}M_\odot{})$ & $R_d$/kpc$^{\rm a}$ & $z_0$\\ 
\hline
A  &  9.6 &  4.43 & 0.1 $R_d$\\
B  &  4.8 &  6.62 & 0.1 $R_d$\\
C  &  4.8 & 13.23 & 0.05 $R_d$\\
\hline
\end{tabular}
\end{center}
\begin{flushleft}
{\footnotesize $^{a}$The values adopted here for $R_d$ are about twice (for runs A and B) and thrice (for run C) the value predicted by Mo, Mao \& White (1998).}\\
\end{flushleft}
\label{tab_1}
\end{table}

As in M07, dark matter particles have mass equal to $4\times{}10^6\,{}M_\odot{}$, whereas disc, bulge and gas particles have mass equal to $4\times{}10^5\,{}M_\odot{}$. Softening lengths are 0.2 kpc for halo particles and 0.1 kpc for disc, bulge and gas particles. The initial smoothing length of the gas is 0.1 kpc.

Thus, the progenitor (intruder) has 122000 (80000) halo dark matter particles. The bulge and the gaseous disc of the progenitor are composed by 60000 and 80000 particles, respectively. The number of stellar disc particles is 240000 in run A, and 120000 in runs B and C.

The three runs differ only in the mass, the scale-length and the scale-height of the stellar disc. The dark matter halo, the bulge, the gaseous disc and the properties of the intruder are identical. In particular, the stellar mass in run A is twice as in runs B and C. The disc scale-length in run C is twice as in run B and a factor of $\sim{}2.8$ larger than in run A. We note that run A is the same as run A3 in M07.

Runs A and B have been integrated for a time $t=1$ Gyr, run C up to $t=1.5$ Gyr.

\section{Ring galaxy evolution}
\begin{figure}
\center{{
\epsfig{figure=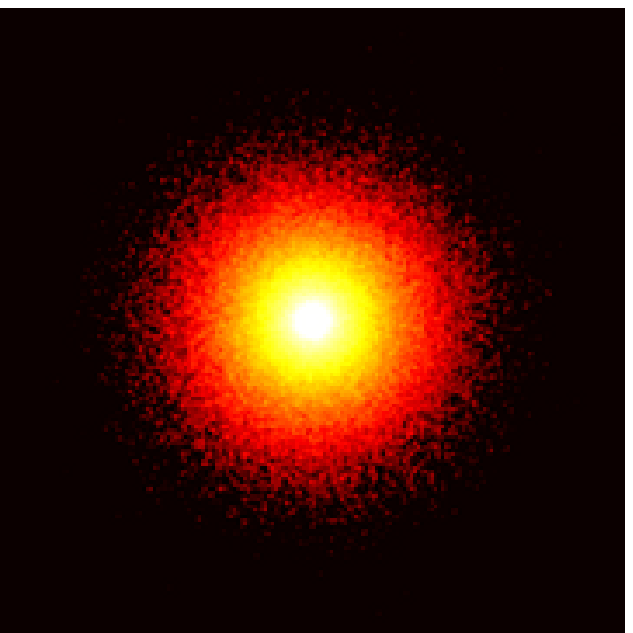,height=4cm} 
\epsfig{figure=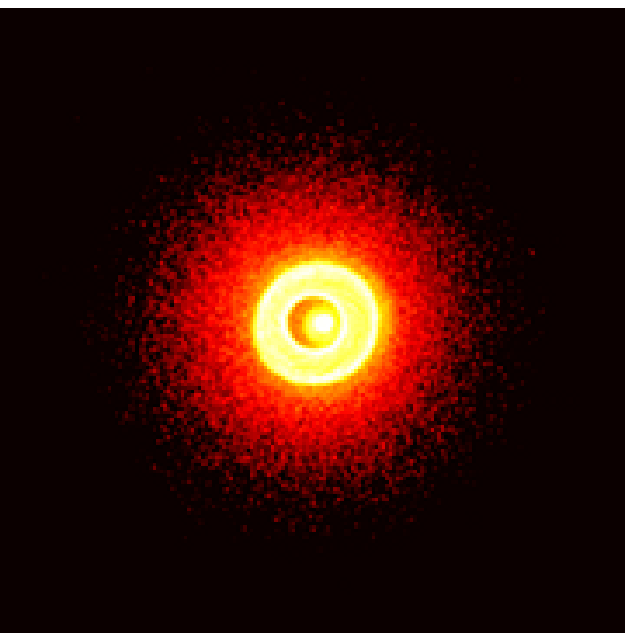,height=4cm} 
\epsfig{figure=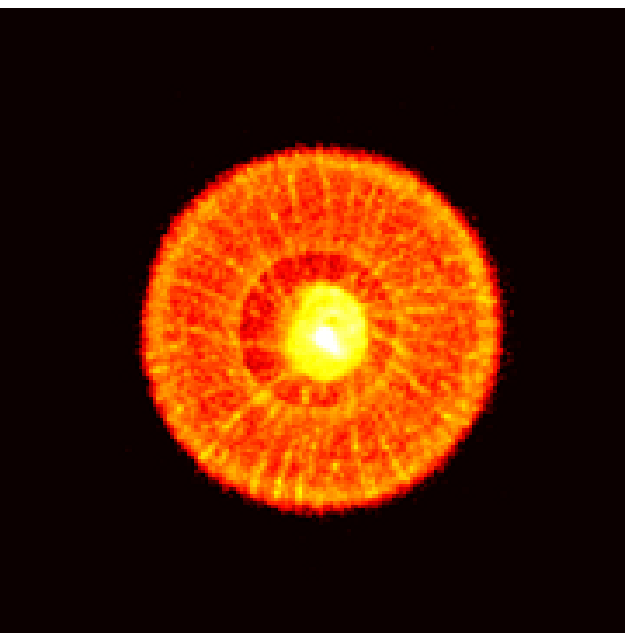,height=4cm} 
\epsfig{figure=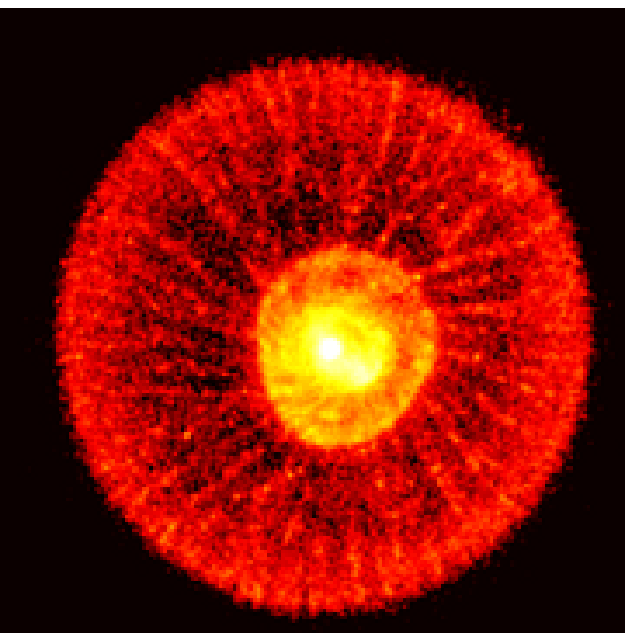,height=4cm} 
}}
\caption{\label{fig:fig1} Time evolution of 
stellar and gas component of the target galaxy (seen face-on) in run C. From top to bottom and from left to right: after 0, 0.16, 0.5 and 1.0 Gyr from the beginning of the simulation. Each frame measures 260 kpc. The colour coding indicates the density, projected along the $z$-axis, in logarithmic scale (from 2 to 70 $M_\odot{}$ pc$^{-2}$).  
}
\end{figure}
\begin{figure}
\center{{
\epsfig{figure=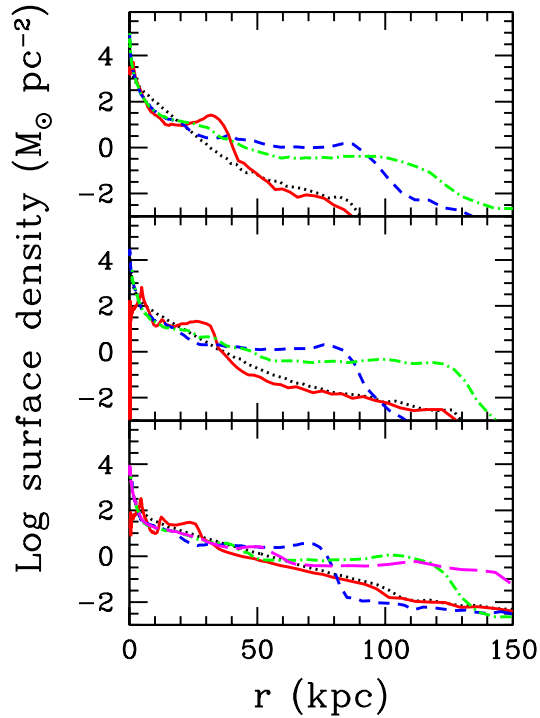,height=10cm} 
}}
\caption{\label{fig:fig2} Stellar surface density in runs A (top panel), B (middle) and C (bottom). In all the three panels the dotted (black on the web), solid (red on the web), short-dashed (blue on the web) and dot-dashed (green on the web) lines correspond to $t=0$, 0.16, 0.5 and 1.0 Gyr, respectively. In the bottom panel the long-dashed line (magenta on the web) corresponds to $t=1.4$ Gyr.}
\end{figure}

\begin{figure}
\center{{
\epsfig{figure=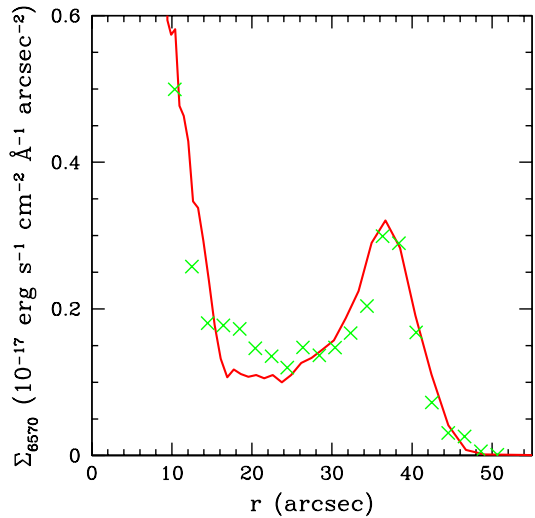,height=8cm} 
}}
\caption{\label{fig:fig3} Comparison between the stellar surface brightness profile of the Cartwheel galaxy and of run A. Crosses (green on the web): observed azimuthally averaged radial surface brightness profile in red continuum of Cartwheel (Higdon 1995). The $1\,{}\sigma{}$ errors are of the same order of magnitude as the crosses. Solid line (red on the web): $R$ magnitude radial surface brightness profile derived from run A at $t=0.16$ Gyr. The simulated profile has been arbitrarily rescaled.}
\end{figure}

Fig.~\ref{fig:fig1} shows the time evolution of a simulated ring galaxy (run C) up to 1 Gyr after the dynamical encounter. The galaxy is seen face-on. At $t=$0.16 Gyr (top right panel) the progenitor has developed a stellar and gaseous ring similar to that observed in ring galaxies (e.g. the Cartwheel galaxy; see Higdon 1995 and references therein). Furthermore the stellar 'spokes', typical of Cartwheel, are well visible in this stage (for an explanation, see M07). Afterward, the ring keeps expanding and fading, while a fraction of the matter inside the ring starts to fall back to the centre. At $t=$0.5 Gyr the ring is  barely visible, while at  $t=$1.0 Gyr it is indistinguishable from the disc. Thus, we conclude that the ring-galaxy phase ends $\approx{}0.5$ Gyr after the interaction.
\begin{figure}
\center{{
\epsfig{figure=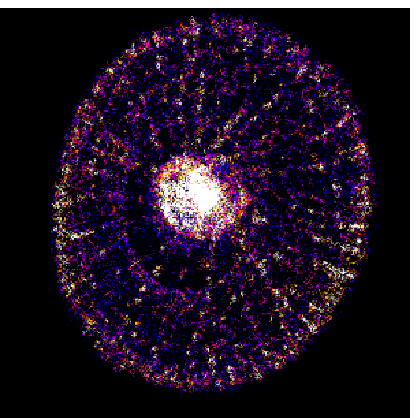,height=4cm} 
\epsfig{figure=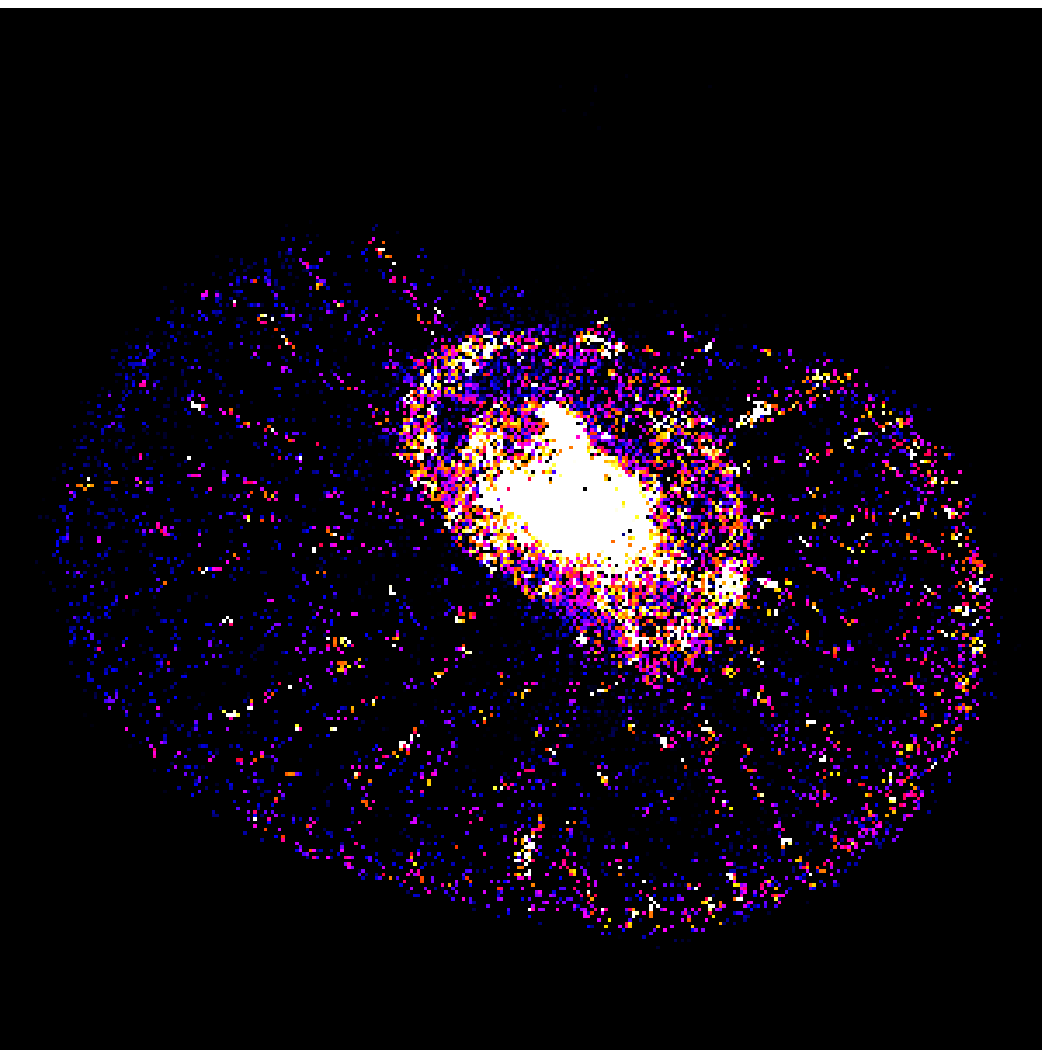,height=4cm} 
\epsfig{figure=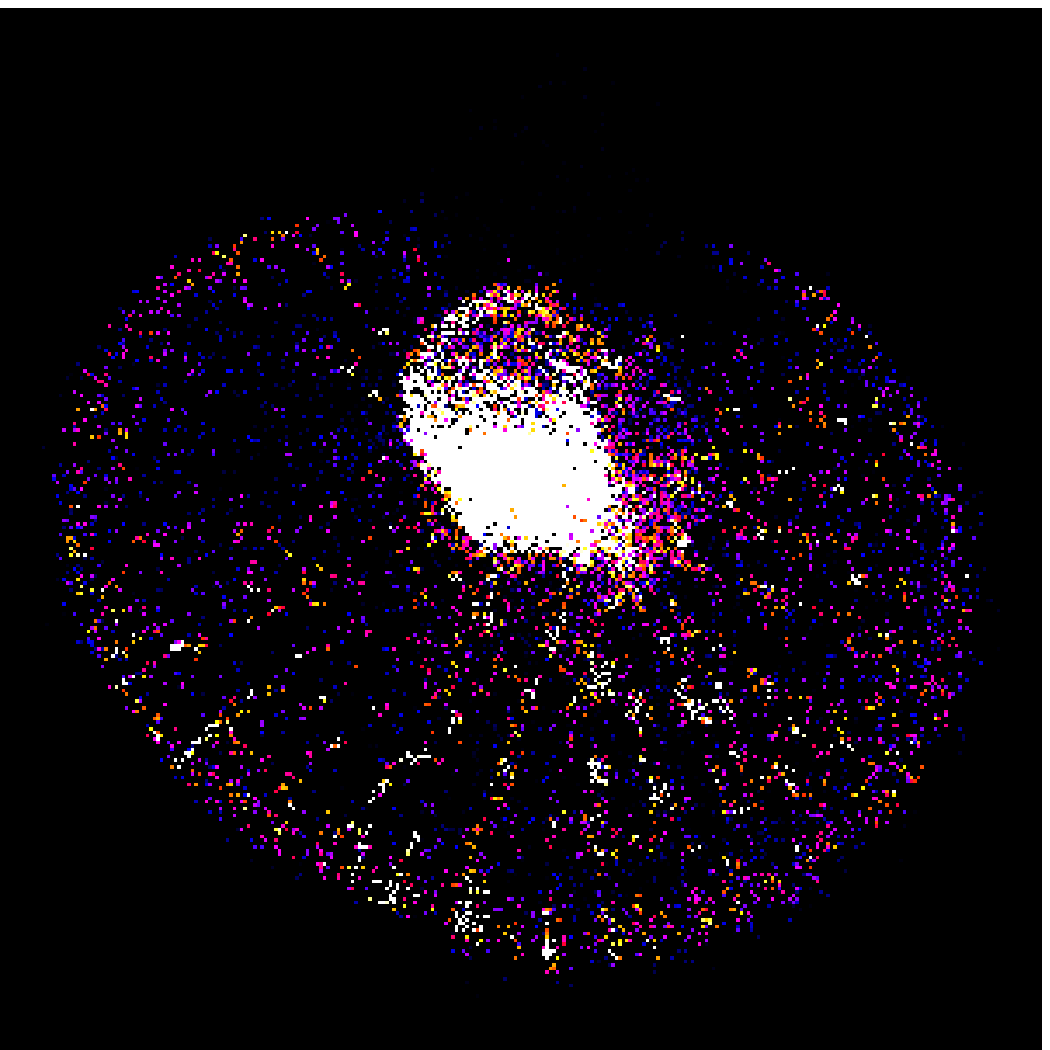,height=4cm} 
\epsfig{figure=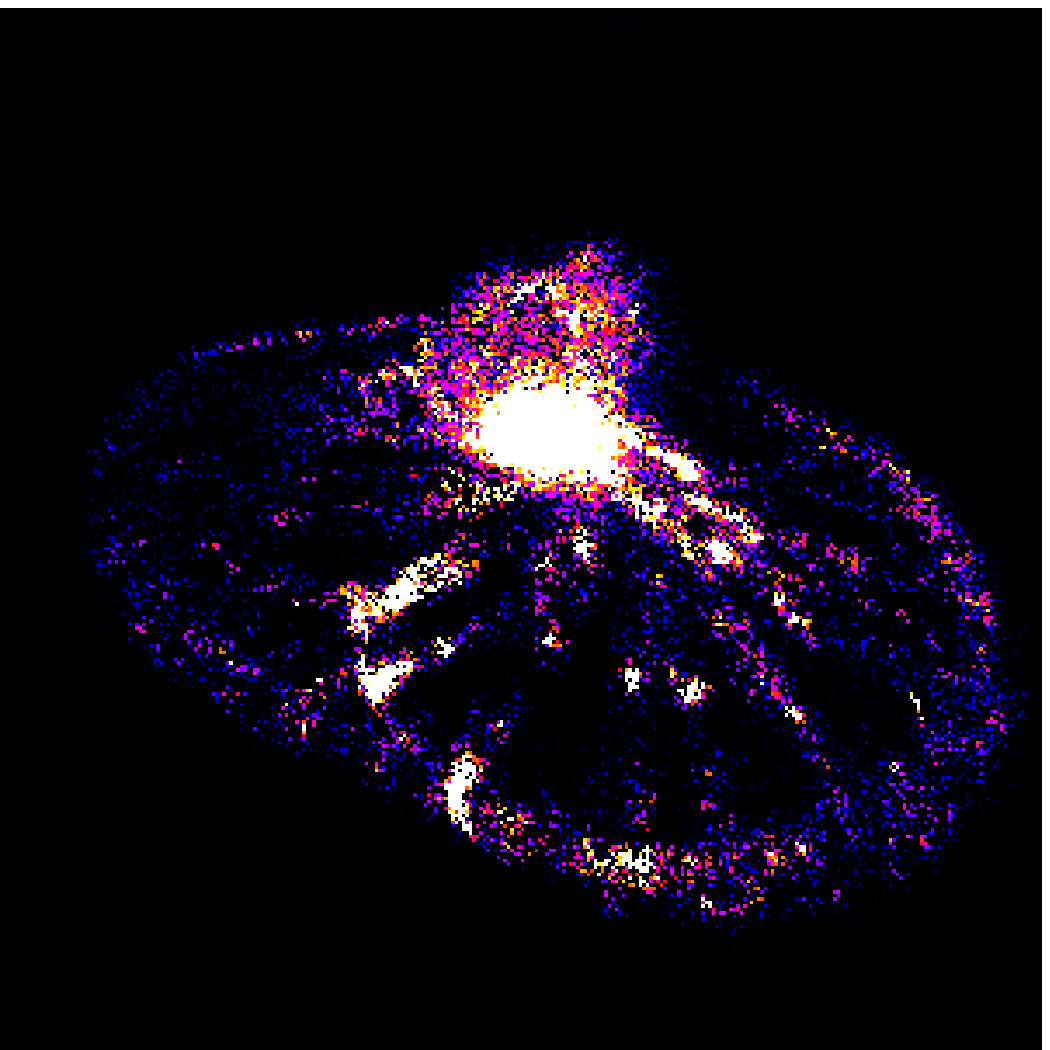,height=4cm} 
}}
\caption{\label{fig:fig4} Stellar density maps for our 4 best matches. From top to bottom and from left to right: run C at $t=0.5$ Gyr with inclination $i=35^\circ$ (which we will compare with UGC6614), run C at $t=1.4$ Gyr with $i=45^\circ{}$ (Malin 1), run B at $t=1.0$ Gyr with $i=38^\circ{}$ (Malin 2), run A at $t=0.5$ Gyr with $i=58^\circ{}$ (NGC7589). From top to bottom and from left to right the frames measure 160, 280, 260 and 180 kpc per edge, respectively. The colour coding indicates the projection along the $z$-axis (i.e. the observer line-of-sight) of the stellar density in linear scale. From top to bottom and from left to right the scale is 0-2.23, 0-0.446, 0-0.223 and 0-2.23 $M_\odot{}$ pc$^{-2}$, respectively.}
\end{figure}

In Fig.~\ref{fig:fig2} the evolution of the stellar surface density is shown, for all the three runs. The initial conditions (dotted line) have a 'standard' bulge$+$exponential profile. At $t=0.16$ Gyr (solid line) a peak is visible at a radius $r\sim{}20-30$ kpc . It corresponds to the propagating outer ring. At this stage our simulation looks like  Cartwheel (see also Fig.~\ref{fig:fig3}). In Fig.~\ref{fig:fig3} the observed surface brightness in red continuum of Cartwheel (crosses) is compared with the surface brightness in $R$-band\footnote{The simulated surface brightness is calculated with the package {\it Tipsy} as described in the next section and arbitrarily rescaled.} derived from run A at $t=0.16$ Gyr. The observations are very well reproduced by the simulation (see also M07).

After $t\approx{}$0.5 Gyr (dashed line in Fig.~\ref{fig:fig2}) the outer ring has propagated up to $\sim{}70-90$ kpc. At this stage, the stellar surface density of the ring has decreased of $\sim{}1$ order of magnitude. On the other hand, the stellar disc has became extraordinarily extended and flat.

At $t\gtrsim{}1$ Gyr the surface density of the ring is $\sim{}2$ orders of magnitude lower than in the 'Cartwheel phase', and is almost comparable with the surrounding density. The stellar disc now extends up to  $100-130$ kpc, showing a flat surface density (in a logarithmic scale).

 The time evolution emerging from our simulations is broadly consistent with the results of analytic theories for ring waves (Struck-Marcell \& Lotan 1990;  Appleton \& Struck-Marcell 1996). For example, in fig.~16 of Appleton \& Struck-Marcell (1996) the strength of the waves at early times suggests that the ring phase is short-lived. In the same figure, at later times, the extended ring wave indicates that the disc has become very large, and the dense phase-mixed central regions confirm that the secondary rings are mixing up in the disc. Then, the time evolution shown by the simulations is the product of the steady development of phase mixing.

\section{Comparison with GLSBs}
The only known discs whose extension and surface density profile are comparable with the late stages of our simulations  are those of the GLSBs. Thus, we decided to investigate this similarity, by comparing our simulations with various observational features of GLSBs (e.g. the surface brightness profile and the rotation curve).

GLSBs are characterised by very low central surface brightness (central $B$ magnitude $\gtrsim{}23$ mag arcsec$^{-2}$). Unlike the other LSBs, GLSBs have unusually large size (up to $100-150$ kpc), high stellar and gas mass ($M_g\gtrsim{}10^{10}\,{}M_\odot{}$), and show a 'normal' stellar bulge.
As they are very extended and diffuse, GLSBs are likely underrepresented in magnitude-limited galaxy surveys (Bothun, Impey \& McGaugh 1997).

At present, $\sim{}18$ galaxies have been classified as GLSBs (Pickering et al. 1997 and references therein). In this paper, we consider 4 of the most massive, best studied, and extended among GLSBs, i.e. Malin 1 (the prototype of the GLSB class and the galaxy with the largest disc scale-length), Malin 2 (also called F568-6), UGC6614 and NGC7589. For further observational details about these 4 galaxies, we refer to Pickering et al (1997) and references therein.

\subsection{Surface brightness profile}
To compare our simulations with the data of these 4 GLSBs, first of all we consider the surface brightness profile. In particular, we compare our 3 runs at different time $t$ with the observed  surface brightness profiles (Pickering et al. 1997; Moore \& Parker 2007) and we find which run (and  which simulation time $t$) best matches the observations. 

\begin{table}
\begin{center}
\caption{List of runs associated with every GLSB.}
\begin{tabular}{cccc}
\hline
\vspace{0.1cm}
GLSB & Run &  $t$ (Gyr) & $i^{\rm a}$ ($^{\circ}$))\\
\hline
UGC6614 & C  & 0.5 & 35\\
Malin 1 & C  & 1.4 & 45\\ 
Malin 2 & B  & 1.0 & 38\\
NGC7589 & A  & 0.5 & 58\\
\hline
\end{tabular}
\end{center}
\begin{flushleft}
{\footnotesize $^{a}$The values adopted here for the inclination are from Pickering et al. (1997).}
\end{flushleft}
\label{tab_2}
\end{table}

Fig.~\ref{fig:fig4} shows the stellar density maps for these 4 best matches.
All these simulated galaxies have been rotated according to the inclination angle $i$ suggested by observations (Pickering et al. 1997). Table 2 lists the details of the best match for each GLSB.

\begin{figure*}
\center{{
\epsfig{figure=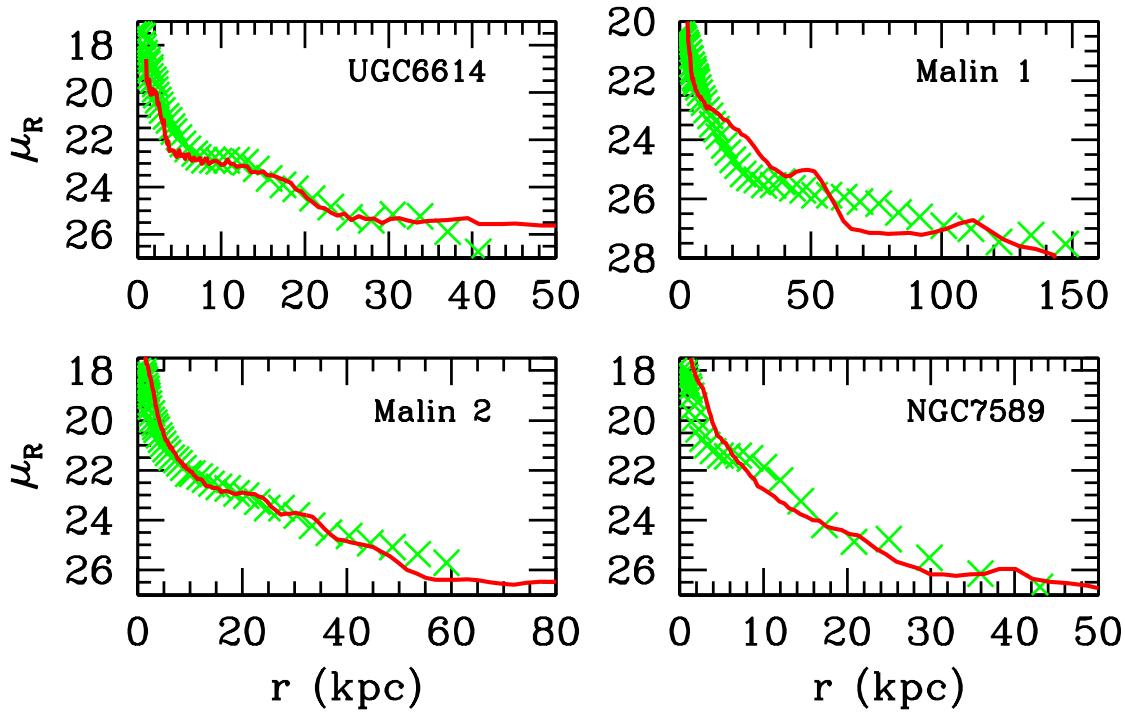,height=10cm} 
}}
\caption{\label{fig:fig5}
$R-$band stellar surface brightness profile of the GLSB sample (in units of magnitude per arcsec$^2$). Crosses (green on the web): data points from Pickering et al. (1997) for UGC6614, Malin 2 and NGC7589, and from Moore \& Parker (2007) for Malin 1. The $1\,{}\sigma{}$ errors are of the same order of magnitude as the points. Solid line (red on the web): stellar surface brightness profile in $R$ magnitude derived from the simulations. From top to bottom and left to right: UGC6614 data and run C ($t=0.5$ Gyr), Malin 1 data and run C ($t=1.4$ Gyr), Malin 2 data and run B ($t=1.0$ Gyr), NGC7589 and run A ($t=0.5$ Gyr). See Table 2. 
 }
\end{figure*}

Fig.~\ref{fig:fig5} shows the comparison among the 4 simulated best matches and the observed stellar surface brightness profile in $R$ magnitude. The data are from Pickering et al. (1997) for UGC6614, Malin 2 and NGC7589, and from Moore \& Parker (2007) for Malin 1.

The $R$ magnitude surface brightness profiles of the simulations have been derived from the stellar surface density (see Fig.~\ref{fig:fig2}) by using the {\it Tipsy} package\footnote{\scriptsize \tt http://www-hpcc.astro.washington.edu/tools/tipsy/tipsy.html}. The method adopted in  {\it Tipsy} for the conversions is described in Katz (1992), and is based on  mass-to-light conversion tables. We upgraded  {\it Tipsy} with respect to Katz (1992), by changing the pre-existing  mass-to-light ratio tables, based on a Salpeter (Salpeter 1955) initial mass function (IMF), with  tables based on a Chabrier IMF (Chabrier 2001). We also updated the conversion from the bolometric magnitude to $V$, $R$ and $B$ magnitude, by using the Padova GALaxies AnD Single StellaR PopulatIon ModELs\footnote{\tt http://dipastro.pd.astro.it/galadriel/} (GALADRIEL; Gilardi et al. 2000). The results are not significantly affected by this change.

From Fig.~\ref{fig:fig5} it appears that the simulations match quite well the observed surface brightness profile. In particular, a $0.5$-Gyr-old ring galaxy (model C) reproduces all the features of the UGC6614 data, included the bump at a distance of $\sim{}20-40$ kpc from the centre. In our simulations this bump has a clear physical meaning: it represents the inner, secondary ring due to the dynamical interaction. Interestingly, $R$-band and H$\alpha{}$ images of UGC6614 (Pickering et al. 1997) show that also in the real galaxy the bump is due to a sort of stellar and gaseous ring-like structure.

The profile of Malin 2  is equally well reproduced by run B at a later stage, i.e. $t=1.0$ Gyr. In fact, this galaxy has a more extended stellar disc with respect to UGC6614, indicating (in our model) that the outer ring had more time to propagate.

The agreement between data and simulations is less good, but still acceptable for Malin 1 and NGC7589. Run~A at $t=0.5$ Gyr can reproduce the profile of NGC7589 (especially the outer part), but does not account for the bump at $\sim{}10$ kpc. Unlike UGC6614, in the case of NGC7589 the bump is too close to the nucleus to be explained with the secondary ring after 0.5 Gyr. It is not unreasonable to suppose that slightly different initial conditions (e.g. the impact parameter and the inclination angle of the intruder with respect to the target) might reproduce also the bump. It is also worth noting that NGC7589 has a gas-rich interacting companion, which could have been the intruder galaxy.

Malin 1 has by far the most extended stellar disc among the known GLSBs. Thus, we have to integrate run C for $1.4$ Gyr, to let the outer ring propagate up to 150 kpc. After $1.4$ Gyr, the simulated disc extends up to $\sim{}140$ kpc; but the surface brightness profile at a distance of $\sim{}20-30$ kpc and  $\sim{}60-80$ kpc is quite different from the observed one. Given the low surface density of star particles at  $\sim{}60-80$ kpc (see Fig.~\ref{fig:fig2}), the discrepancy at these radii is probably due to density fluctuations. The discrepancy at $\sim{}20-30$ kpc is more serious, as it happens in a higher density region. 
Nevertheless, it is the first time that a disc with the same extension as that of Malin 1 is reproduced  by a simulation started from initial conditions consistent with the CDM scenario.

\subsection{Morphological features of GLSBs}
All the GLSBs are characterised by the presence of a bulge and by the unusual extension, the low surface brightness and the flat slope of the stellar disc. As we already discussed, these features are well reproduced by our simulations. Furthermore, there are also other (less evident) morphological features which characterise a fraction of GLSBs (e.g. the presence of a bar). In this section, we briefly consider the most intriguing among these characteristics and we compare them with our simulations.

\begin{figure}
\center{{
\epsfig{figure=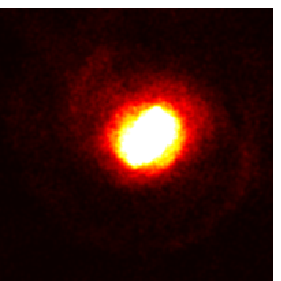,height=4cm} 
\epsfig{figure=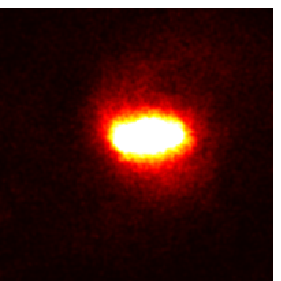,height=4cm} 
\epsfig{figure=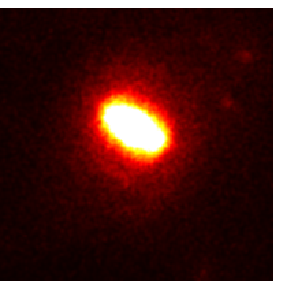,height=4cm} 
\epsfig{figure=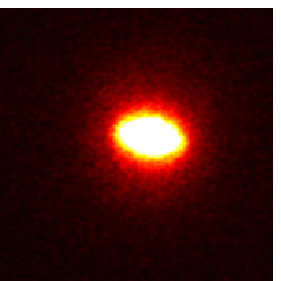,height=4cm}     
}}
\caption{\label{fig:figboh3} Development of a bar in run A. From top to bottom and from left to right: $t=0.2$, 0.3, 0.5, 1.0 Gyr. Each frame measures 25 kpc per edge. The colour coding indicates the projection along the $z$-axis (i.e. the observer line-of-sight) of the stellar density in linear scale. The scale is 0-1000 $M_\odot{}$ pc$^{-2}$.
}
\end{figure}
Some GLSBs (e.g. Malin 1 and NGC7589) present  a stellar bar. In the case of Malin 1 the length of the bar is $\sim{}8-9$ kpc (Barth 2007). In our simulations we can also observe the formation of a bar, which is triggered by the galaxy interaction. The bar develops after the first 200$-$300 Myr from the beginning of the simulation (Fig.~\ref{fig:figboh3}), i.e. $\sim{}150-250$ Myr after the galaxy collision, and  is quite long-lived, as it is still evident at $t=1$ Gyr. The length of the bar is $\sim{}8-10$ kpc, consistent with the observed bar in Malin~1.

The formation and the final size of the bar in our simulations strongly depend on the stellar disc mass $M_d$: in run A, which has twice the disc mass as run B and C, the bar is much more evident than in the other two runs. 
 
Furthermore, some GLSBs show faint and disturbed spiral arms (e.g. Malin 1). In our simulations we do not observe evident spiral arms, but this might be due to a numerical resolution problem. Another possibility (especially for Malin 1 and UGC6614) is that the secondary ring itself (deformed by disc rotation) can be observationally confused with a spiral arm.

\begin{figure}
\center{{
\epsfig{figure=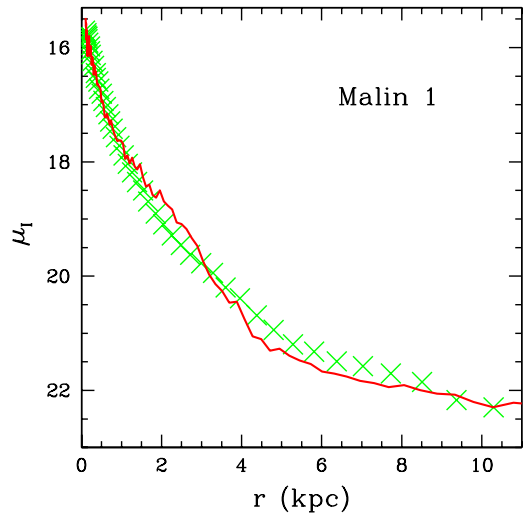,height=8cm} 
}}
\caption{\label{fig:figboh4} 
$I-$band stellar surface brightness profile of the central part of Malin 1 (in units of magnitude per arcsec$^2$). Crosses (green on the web): data points from Barth (2007). The $1\,{}\sigma{}$ errors are of the same order of magnitude as the points. Solid line (red on the web): stellar surface brightness profile in $I$ magnitude derived from the simulation (run C at $t=1.4$ Gyr). 
}
\end{figure}

Finally, Barth (2007) claimed the existence of a normal exponential disc in the central  $\lesssim{}9$ kpc of Malin 1. This exponential disc is very small with respect to the flat extended disc of Malin 1 and is difficult to disentangle from the bar component. In Fig.~\ref{fig:figboh4} the $I$ magnitude data from Barth (2007) are compared with the  profile extracted from our simulations (run C at $t=1.4$ Gyr). Our run reproduces quite well the central part ($\lesssim{}10$ kpc) of Malin 1's profile: the bulge, the bar and even the exponential disc component, visible in the observations, are evident also in the simulated profile.  

\subsection{Colours and star formation}
\begin{figure}
\center{{
\epsfig{figure=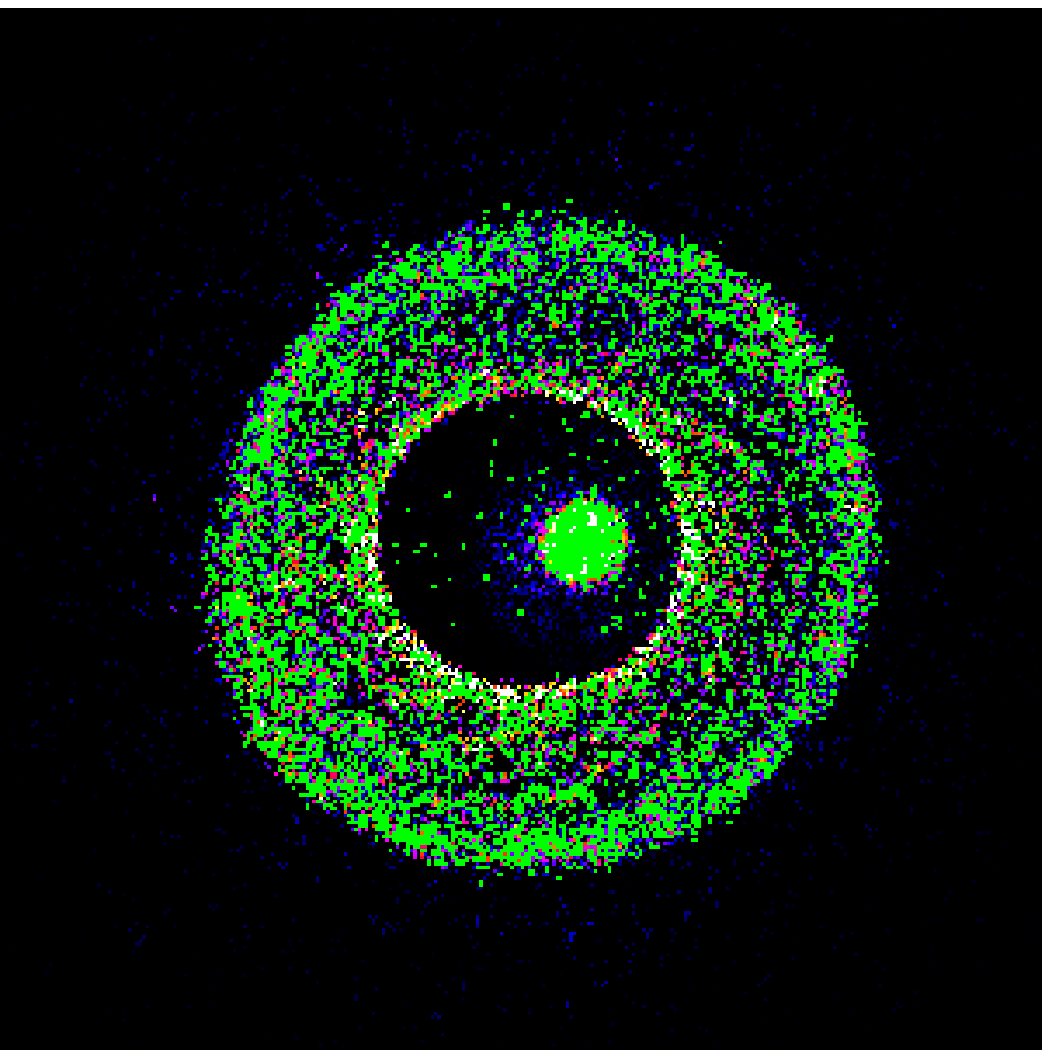,height=4cm} 
\epsfig{figure=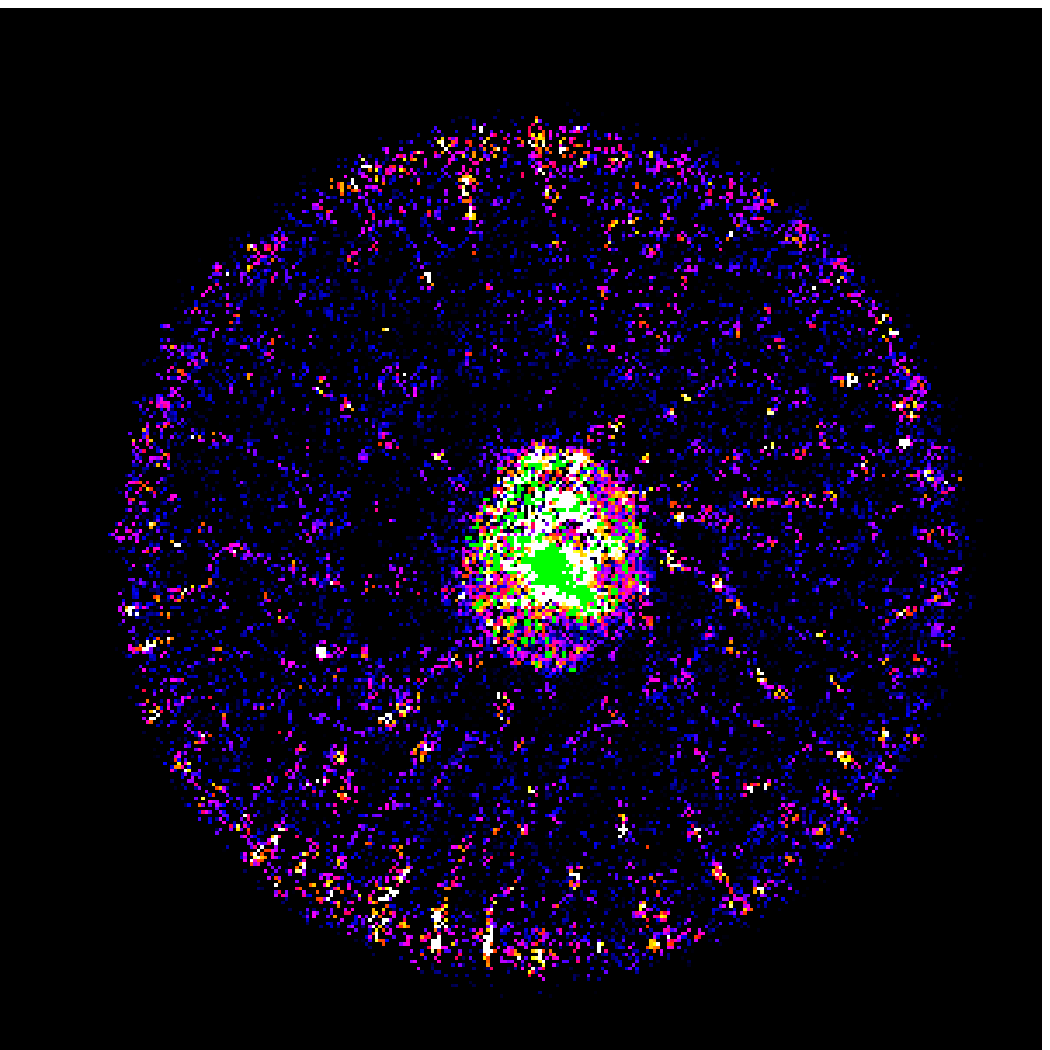,height=4cm} 
\epsfig{figure=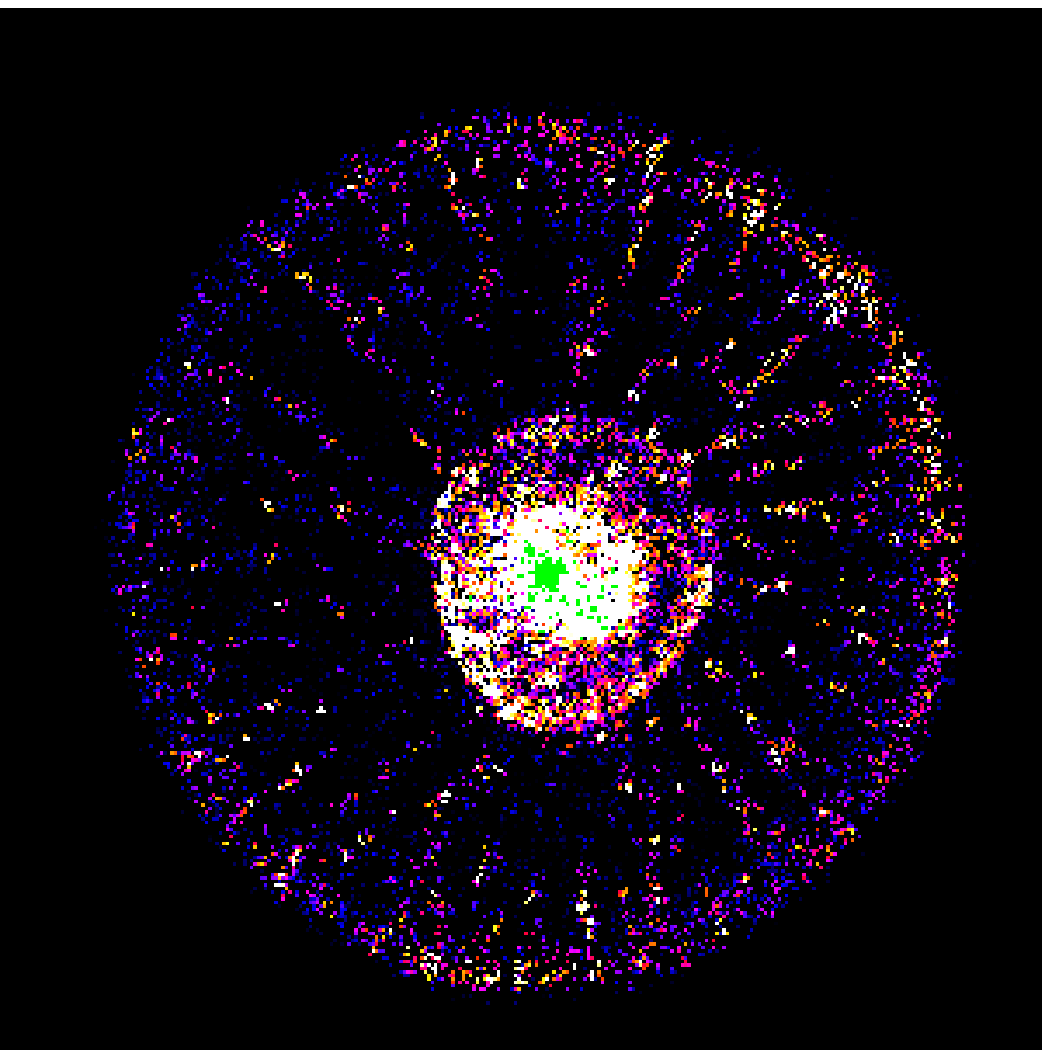,height=4cm} 
\epsfig{figure=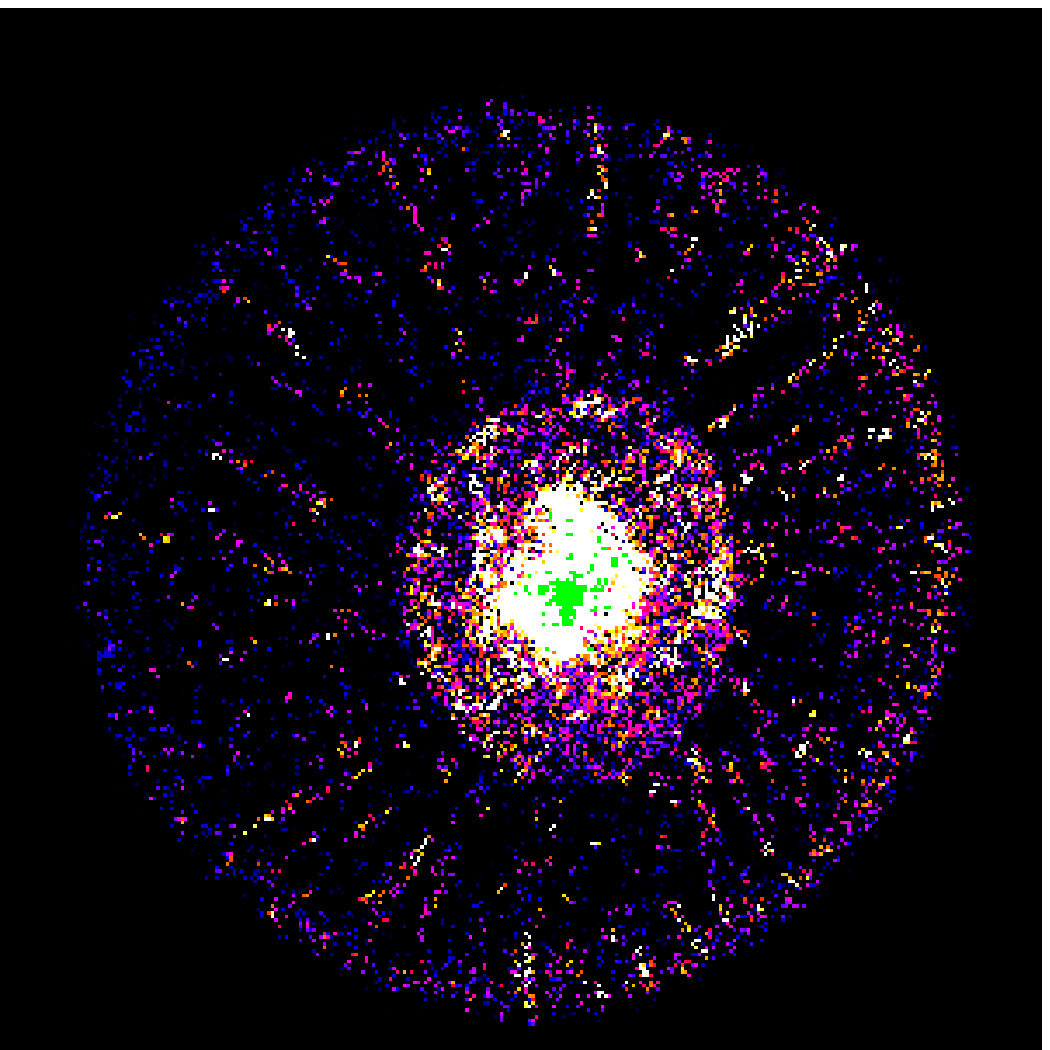,height=4cm} 
}}
\caption{\label{fig:figboh} Young stellar particles (filled circles, green in the online version) superimposed to the stellar density maps for run C.
 Young stellar particles are $\leq{}$100 Myr old. From top to bottom and from left to right: $t=0.16$, 0.5, 1.0, 1.4 Gyr. Top and bottom right panels match UGC6614 and Malin 1, respectively.
From top to bottom and from left to right the frames measure 80, 180, 270 and 300 kpc per edge, respectively. The colour coding indicates the projection along the $z$-axis (i.e. the observer line-of-sight) of the stellar density in linear scale. From top to bottom and from left to right the scale is 0-44.6, 0-4.46, 0-0.892 and 0-0.446 $M_\odot{}$ pc$^{-2}$, respectively.
}
\end{figure}

\begin{figure}
\center{{
\epsfig{figure=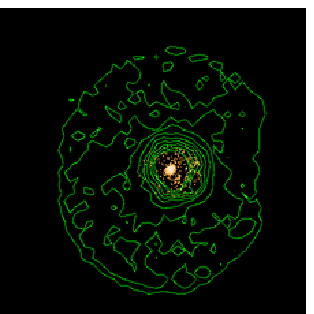,height=4cm} 
\epsfig{figure=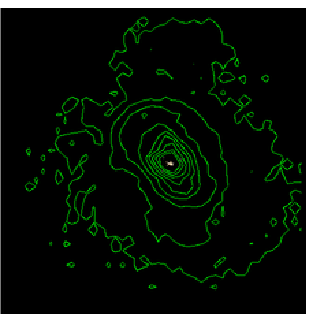,height=4cm} 
\epsfig{figure=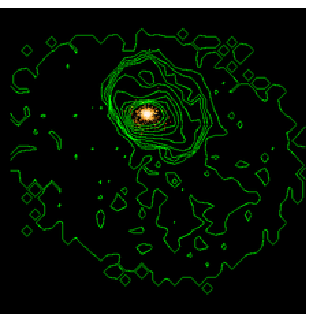,height=4cm}
\epsfig{figure=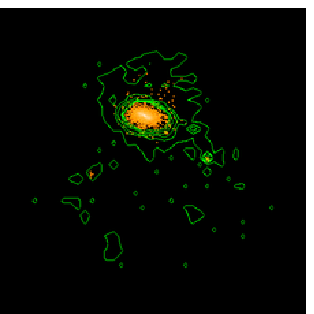,height=4cm}
}}
\caption{\label{fig:figboh2} Isocontours of the gas surface density superimposed to the H$\alpha{}$ emission map associated with young stars (defined as in Fig.~\ref{fig:figboh}) for the 4 best matches (see Fig.~\ref{fig:fig4} and Table 2). From top to bottom and from left to right: best match for UGC6614, Malin 1, Malin 2, NGC7589. Contour levels range from the peak density to $\sim{}$1 per cent of the peak density and are logarithmically scaled. The simulated  H$\alpha{}$ emission maps scale logarithmically from the peak intensity (white in the Figure) to 10 per cent (gray, orange in the web) of the peak intensity. The total H$\alpha{}$ luminosity derived from the simulations is $1.9\times{}10^{41}$, $6.6\times{}10^{40}$, $3.3\times{}10^{41}$ and $1.5\times{}10^{42}$ erg s$^{-1}$ (corresponding to a SFR of 1, 0.3, 1.8 and 8.1 $M_\odot$ yr$^{-1}$) for UGC6614, Malin 1, Malin 2 and NGC7589, respectively. 
}
\end{figure}
Not only the surface brightness profile, but also the colour and the colour gradient of GLSBs are interesting features to compare with our simulations. The observed GLSBs are considerably redder than LSBs (Sprayberry et al. 1995). The colours and spectroscopic indexes of the bulges of GLSBs are indistinguishable from those of normal spiral galaxies (Sprayberry et al. 1995). However, also disc stars are quite red, indicating, together with the weakness of H$\alpha{}$ emission, that star formation is no longer active in the discs of GLSBs (Pickering et al. 1997). The only available estimate of the H$\alpha{}$ luminosity, $L_{H\alpha{}}$, (Impey \& Bothun 1989) indicates that $L_{H\alpha{}}\approx{}10^{40}$ erg s$^{-1}$ for Malin 1, corresponding to a star formation rate (SFR) of $\approx{}0.1\,{}M_\odot{}$ yr$^{-1}$.

All these characteristics of GLSBs match quite well our simulations. At the stage in which  our simulations are comparable with GLSB surface brightness, most of simulated stars  are quite old: in run C, at $t=1.4$ Gyr $\sim{}97$ per cent of stars are more than 1-Gyr old. 
Most of stars already exist in the initial conditions (and are assumed to be 3-Gyr old at $t=0$).
The bulk of the newly formed bulge and disc stars is born within the first 200 Myr (Fig.~\ref{fig:figboh}) from the beginning of the simulation (i.e. immediately after the collision).
In fact, after a burst of star formation between 40 and 200 Myr (as a consequence of the galaxy collision, see M07 for details), at  $t\gtrsim{}500$ Myr the gas density in the disc has become so low that no significant star formation is ongoing outside the bulge (Fig.~\ref{fig:figboh}).

Fig.~\ref{fig:figboh2} shows the simulated H$\alpha{}$ emission associated with young stars. In all the 4 galaxies the H$\alpha{}$ emission predominantly comes from the bulge. Malin 1 and Malin 2 have only a very concentrated H$\alpha{}$ spot. The two younger galaxies, UGC6614 and NGC7589, have also a more diffuse component. In UGC6614 this external component presents a ring-like shape (corresponding to the inner ring), while in NGC7589 the H$\alpha{}$ emission comes from a sort of bar. These features agree with the observed H$\alpha{}$ maps shown in fig.~4 of Pickering et al. (1997). The total  H$\alpha{}$ luminosity derived from the simulations is $1.9\times{}10^{41}$, $6.6\times{}10^{40}$, $3.3\times{}10^{41}$ and $1.5\times{}10^{42}$ erg s$^{-1}$ (corresponding to a SFR of 1, 0.3, 1.8 and 8.1 $M_\odot$ yr$^{-1}$) for UGC6614, Malin 1, Malin 2 and NGC7589, respectively. The value obtained for  Malin 1 is approximately in agreement with Impey \& Bothun (1989). The SFR derived for NGC7589 ($8.1\,{}M_\odot$ yr$^{-1}$), significantly higher than that of the other 3 galaxies, is also consistent with the observed H$\alpha{}$ maps and with the brightness of NGC7589 in the ultraviolet band ($L_{\rm UV}\sim{}10^{43}$ erg s$^{-1}$ both at 1516 \AA{} and at  2267 \AA{}; Gil de Paz et al. 2007). 

Fig.~\ref{fig:figboh2} also shows the density contour levels of atomic hydrogen gas particles (i.e. gas particles with temperature $\sim{}20000$ K and density $\lesssim{}10\,{}$cm$^{-3}$). These contours can be qualitatively compared with the HI intensity contours in fig.~4 of Pickering et al. (1997). The agreement is quite good, especially for UGC6614. From Fig.~\ref{fig:figboh2} the gaseous disc appears extremely extended and its surface density is generally low, similarly to the stellar disc.

 In summary, all our simulated galaxies appear to be in a 'quiescent' stage, subsequent to the starburst phase which was triggered by the galaxy collision and lasted $\sim{}200-300$ Myr. In the GLSB stage, the gaseous disc has expanded so much that, although the total amount of gas is still large ($\gtrsim{}10^{10}\,{}M_\odot{}$), the local gas density is too low to permit star formation. Only NGC7589 still has a relatively high star formation rate (both in the observations and in the simulations). Interestingly, NGC7589 is the only galaxy of our sample which still shows a very close, possibly interacting companion (F893-29, Pickering et al. 1997).

\subsection{HI spectra}\label{sec:spectra}
Pickering et al. (1997) also report Very Large Array (VLA) spectra of the HI in  these 4 GLSBs. To compare our results with these data, we have rotated the simulations of the inclination angle (see Table 2) adopted in Pickering et al. (1997). Then, we calculated the gas velocity along the line-of-sight.

\begin{figure}
\center{{
\epsfig{figure=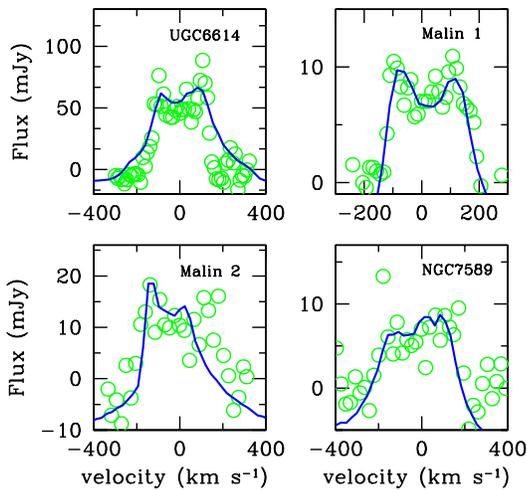,height=8cm} 
}}
\caption{\label{fig:fig6} 
HI spectra of the GLSB sample. Open circles 
 (green on the web): data from Pickering et al. (1997). 
Solid line (blue on the web): spectra derived from our simulations and arbitrarily rescaled. From top to bottom and from left to right: UGC6614, Malin 1, Malin 2 and NGC7589.}
\end{figure}

The results are shown in Fig.~\ref{fig:fig6}. Even in this case there is good agreement between data and simulations, especially for UGC6614 and Malin 1, which clearly show the double-horned spectrum observed in all the GLSBs.
 
\subsection{Rotation curves}
Pickering et al. (1997) also report the HI rotation curves for the GLSB sample.  We extract the gas rotation curves from our simulations, by adopting a procedure as similar as possible  to the one used by Pickering et al. (1997). In particular, after rotating the simulated galaxy (according to the observed inclination angle), we divide  it in concentric annuli and we calculate the gas line-of-sight velocity in each annulus.
\begin{figure}
\center{{
\epsfig{figure=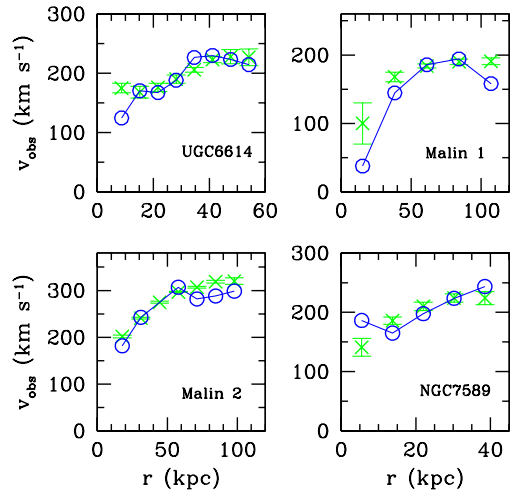,height=8cm} 
}}
\caption{\label{fig:fig7} 
HI rotation curves of the GLSB sample. Crosses (green on the web) are observational data from Pickering et al. (1997). $1\,{}\sigma{}$ errors are shown. Open circles connected by the solid line (blue on the web) are the simulations. From top to bottom and from left to right: UGC6614 (run C at $t=0.5$ Gyr), Malin 1 (run C at $t=1.4$ Gyr), Malin 2 (run B at $t=1.0$ Gyr), NGC7589 (run A at $t=0.5$ Gyr). The simulation B has been rescaled to match Malin 2.}
\end{figure}

The comparison with the data is shown in Fig.~\ref{fig:fig7}. In the case of UGC6614 the agreement between simulations and observations is quite good. In general, the simulated rotation curves do not match the data as well as in the case of the surface brightness profile. However, we point out that the error bars in Pickering et al. (1997) are probably underestimated, as the residuals to their model are quite large (see fig.~8 of Pickering et al. 1997). 

Recently, Sancisi \& Fraternali (2007) have re-analysed the HI data of Malin 1 from Pickering et al. (1997), adopting a different method to derive the rotation curve\footnote{Pickering et al. (1997) derived the rotation curve from the intensity-weighted mean velocities. Sancisi \& Fraternali (2007) have taken the velocities at the profile peaks.}. The rotation curve by Sancisi \& Fraternali (2007)  differs from the one reported by Pickering et al. (1997) and from our simulations only for the rise in the central part ($\lesssim{}30$ kpc).
We stress that in analysing our simulation we used a method analogous to the one adopted by Pickering et al. (1997). Thus, it is natural that our rotation curve is more similar to that in Pickering et al. (1997) than in Sancisi \& Fraternali (2007). 
Nevertheless, the new results by Sancisi \& Fraternali (2007) are  further evidence that the rotation curve of Malin 1 strongly depends on the adopted method of data analysis, and is affected by a larger uncertainty than indicated by the error bars.

 We also stress that the simulated rotation curve that we obtain from our method is different from the circular velocity, which depends only on the enclosed mass and which, in our simulations, slightly increases towards the centre. This difference is due to the fact that, using the procedure described above, we have the same problem as the observers have, i.e. we consider the entire line-of-sight velocity and we cannot disentangle the circular motions from the non-circular ones.

\begin{figure}
\center{{
\epsfig{figure=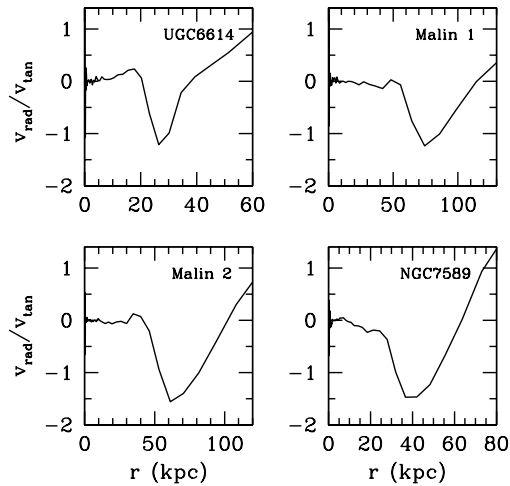,height=8cm} 
}}
\caption{\label{fig:fig12} 
Ratio between radial ($v_{\rm rad}$) and tangential velocity ($v_{\rm tan}$) of gas in the simulations. From top to bottom and from left to right: UGC6614 (run C at $t=0.5$ Gyr), Malin 1 (run C at $t=1.4$ Gyr), Malin 2 (run B at $t=1.0$ Gyr) and NGC7589 (run A at $t=0.5$ Gyr).}
\end{figure}

Furthermore, in all these GLSBs (with the possible exception of UGC6614), the gas appears strongly warped and irregular (Pickering et al. 1997), indicating the existence of strong non-circular motions. Such circumstance might favour our scenario, as the non-circular motions could be associated with the expansion velocity of the old ring and/or with the fallback of a part of the ejected matter toward the centre. From our simulations we can also quantify the strength of non-circular motions. Fig.~\ref{fig:fig12} shows the ratio  between radial ($v_{\rm rad}$) and tangential velocity ($v_{\rm tan}$) of gas in our simulations. In the
inner part of the four galaxies, $v_{\rm rad}$ is negligible, if compared with $v_{\rm tan}$. In the outer part, $v_{\rm rad}$ is comparable with $v_{\rm tan}$. In particular, there is a region where radial velocities are strongly negative (as the gas is falling back), followed by a more external region where the ring is still expanding. It would be interesting to check whether observations confirm this result.


\section{Conclusions}
In this paper we presented a numerical model of ring galaxy evolution. About $100-200$ Myr after the collision with the intruder, the target disc galaxy evolves into a ring galaxy, similar to Cartwheel. Afterward, the ring progressively expands and fades. After $\approx{}0.5-1.0$ Gyr the ring is no longer distinguishable from the disc, its surface density is more than 1 order of magnitude lower than in the Cartwheel phase, and the disc extends up to $\sim{}100$ kpc.

We showed that simulated ring galaxies in the late stages of their dynamical evolution ($\gtrsim{}500$ Myr) are very similar to the observed GLSBs. 
The $R$-band surface brightness profile, the SFR, the HI spectra and also the rotation curves of four GLSBs are well reproduced by the simulations. This result is unlikely due to a simple coincidence.

If all GLSBs were originated by the evolution of P-type ring galaxies, their current number density would be comparable to the observed number density of P-type ring galaxies, i.e. $\sim{}5.4\times{}10^{-6}\,{}h^3\,{}{\rm Mpc}^{-3}$ (where $h$ is the Hubble constant; Few \& Madore 1986). 
 Since we know $\sim{}17$ galaxies which can be classified as GLSBs (Sprayberry et al. 1995; Pickering et al. 1997) and which are within $\sim{}340$ Mpc from our Galaxy, the lower limit of the current number density of GLSBs is  $\sim{}2.7\times{}10^{-7}\,{}h^3\,{}{\rm Mpc}^{-3}$, i.e. about one order of magnitude lower than the density of ring galaxies. This can imply either that a large fraction of GLSBs have not been detected yet, or that only the $\sim{}$5 per cent of ring galaxies ends up into a GLSB. The former scenario is quite realistic, as magnitude-limited surveys are strongly biased against GLSBs (Bothun et al. 1997). The latter hypothesis is also likely, as ring galaxies can end their life in other ways. For example, if the intruder is not sufficiently massive, the density wave is not strong enough to produce a GLSB, or, if the relative velocity is not sufficiently high, the companion can come back and merge or disrupt the propagating wave. Furthermore, recycled dwarf galaxies might also form from the debris of old collisional ring galaxies (e.g. the case of NGC~5291; Bournaud et al. 2007).

Furthermore, our model can explain the formation of GLSBs within the context of the CDM scenario,  in which very extended discs are hardly explained
as a result of normal hierarchical assembly.
Hence it will be quite important to check the consistency of our model with future observations. For example, the available HI data (Pickering et al. 1997) show that most of GLSBs have strong non-circular motions. From the published data it is not possible to understand whether these motions are consistent with the expansion and/or the falling back of gas in the ring. Thus, it is very important to make new radio observations of GLSBs or, at least, re-examine the archival data.

Another important point to address is how many GLSBs have a nearby companion which could be the  intruder. Among the 4 GLSBs considered here, NGC~7589 has a well-known interacting companion (Pickering et al. 1997). Furthermore, UGC~6614 is  surrounded by other galaxies with approximately the same redshift (e.g. KUG 1136+173, AGC~211143 and CGCG~097$-$034), but no studies have been done to establish whether they are in the same group as UGC6614. Finally, the surroundings of Malin~1 are populated by many companion candidates, but no redshift measurements are available at present (Sprayberry et al. 1995).

 A further issue raised by this paper is whether there are objects in the intermediate stage between ring galaxies and GLSBs. The ring galaxy Arp~10 has been thought to be a relatively old ring galaxy, because its rings are not as prominent as those of other ring galaxies (e.g. Cartwheel) and because its SFR, from H$\alpha{}$ observations, appears quite low (Charmandaris, Appleton \& Marston 1993). However, Bizyaev, Moiseev \& Vorobyov (2007) have shown that the SFR of Arp~10, from far-infrared observations
, is $\sim{}10-21\,{}M_\odot{}$ yr$^{-1}$ (analogous to the one of Cartwheel; Mayya et al. 2005), and that the collision which produced the ring occurred only $\sim{}85$ Myr ago. 
Then, both current observational data and theoretical models are not sufficient to support the idea that Arp~10 is an old ring galaxy\footnote{Even our simulations cannot solve the problem, as they indicate that the surface brightness profile of Arp~10 can be matched by a $\sim{}80$ Myr old ring galaxy (in agreement with Bizyaev et al. 2007), as well as by a $\sim{}300$ Myr old ring galaxy. In particular, the surface brightness profile of Arp~10 is matched by a $80$ Myr old ring galaxy with the same initial conditions as run A but with $R_d=8.8$ kpc. Nice agreement with the observations is obtained also for a $300$ Myr old ring galaxy with the same initial conditions as runs B in table 1 of M07, but with $R_d=13.2$ kpc.}.

On the other hand, one of the four GLSBs considered in this paper, UGC6614, shows (both in H$\alpha{}$, in HI and in optical) structures which look like the remnants of the inner ring. Thus, UGC6614 could be in an intermediate, connecting stage between GLSBs and ring galaxies. It would be crucial to study more deeply UGC6614, as well as to search for other galaxies which could represent the intermediate phase between GLSBs and ring galaxies.

\section*{Acknowledgments}
The authors thank the referee, Curt Struck, for his helpful comments.
The authors also thank R.~Sancisi, S.~Courteau and G.~Lake for useful discussions, and acknowledge S.~Callegari, P.~Englmaier, J.~Stadel and D.~Potter for technical support.
MM acknowledges support from the Swiss
National Science Foundation, project number 200020-109581/1
(Computational Cosmology \&{} Astrophysics).

{}

\end{document}